\documentclass[aps,prl,twocolumn,superscriptaddress]{revtex4-2}

\usepackage{amsmath,amssymb,bm}
\usepackage{graphicx}
\usepackage{hyperref}
\begin{document}

\title{Efficient spin--lattice dynamics and spin thermometry with machine-learned magnetic potentials}

\author{Yuanmai Zheng}
\author{Liangliang Hong}
\author{Xingao Gong}
\author{Hongjun Xiang}
\email{hxiang@fudan.edu.cn}
\author{Hongyu Yu}
\email{hongyuyu20@fudan.edu.cn}
\affiliation{Key Laboratory of Computational Physical Sciences (Ministry of Education), Institute of Computational Physical Sciences, State Key Laboratory of Surface Physics, and Department of Physics, Fudan University, Shanghai 200433, China}

\date{\today}

\begin{abstract}
In finite-temperature spin--lattice dynamics, conventional force-only spin thermometers fail for energies nonlinear in each spin, while the standard semi-implicit SIB integrator requires two model evaluations per step. Here we introduce a geometric-probe spin temperature that is exact for arbitrary high-order smooth Hamiltonians using first-order spin forces alone without Hessians, making it naturally suited for nonlinear many-body machine-learning (ML) magnetic potentials. Complementarily, our cached-field integrator (CFI) stays second-order accurate with a single model evaluation per step, about halving the time per step with ML potentials. Extensive benchmarks against Monte Carlo across analytic and ML systems confirm both exact thermometry and accurate thermodynamics.
\end{abstract}

\maketitle

Magnetic materials couple spin and lattice degrees of freedom, and this coupling controls ultrafast demagnetization \cite{Beaurepaire1996,Chen2025}, magnon--phonon hybridization \cite{Man2017}, thermal transport \cite{Zhou2020}, the temperature dependence of magnetic anisotropy and magnetoelasticity \cite{Nikolov2023,Nikolov2021}, and the Einstein--de Haas effect \cite{Dednam2022}.
Accurate finite-temperature simulation of magnetic systems therefore requires spin--lattice dynamics (SLD), which evolves atomic coordinates and classical spins together \cite{Ma2008,Ma2012,Tranchida2018,Hellsvik2019,Strungaru2021} by coupling molecular dynamics (MD) with the stochastic Landau--Lifshitz--Gilbert (LLG) equation and Langevin baths \cite{Brown1963,GarciaPalacios1998,Ma2011}.
Magnetic machine-learned (ML) potentials now provide near-first-principles energies and forces for such simulations \cite{Eckhoff2021,Domina2022,Novikov2022,Yu2022,Yu2024,Yu2024b,Yang2024,Yuan2024,Kotykhov2023,Chapman2022,Rinaldi2024}, for example for bcc Fe \cite{Novikov2022,Chapman2022,Rinaldi2024,Brannvall2022}, Fe--Al alloys \cite{Kotykhov2023}, Fe$_3$GeTe$_2$ \cite{Yu2022}, monolayer CrI$_3$ and CrTe$_2$ \cite{Yu2024b}, BiFeO$_3$ \cite{Cheng2026,Zhang2026}, and UO$_2$ \cite{Kobayashi2026}.

Two problems limit quantitative SLD with these potentials.
First, when the spin and lattice baths are controlled separately, the spin temperature must be measured to check that the spins reach their bath temperature, because finite time steps can shift it; LLG dynamics, however, has no spin momenta from which to read it.
The dynamical spin temperature of Nurdin and Schotte \cite{Nurdin2000,Ma2010}, in the simplified form reported by the LAMMPS SPIN package \cite{Tranchida2018}, uses only spin orientations and spin forces but is exact only when the Hamiltonian $H$ is linear in each spin; single-ion anisotropy, biquadratic exchange, and ML many-body terms violate this condition.
Generalized configurational temperatures \cite{Rugh1997,Butler1998,Jepps2000,Davis2012,Palma2016,Gutierrez2018} hold for any smooth $H$, but the estimators used in practice, of Nurdin--Schotte or Rickayzen type \cite{Nurdin2000,Rickayzen2001,Gutierrez2018}, need second derivatives of $H$; for a magnetic ML potential these require second-order automatic differentiation through the network, which multiplies the memory use and the evaluation time.
Second, model evaluations dominate the cost of an SLD step with an ML potential: the semi-implicit SIB integrator \cite{Mentink2010}, used in SLD codes \cite{Muller2019,Hellsvik2019}, calls the model twice per step, once for the spin update and once for the atomic forces.

Earlier work improved norm preservation with Cayley and geometric maps \cite{Krishnaprasad2001,Frank2004,Dahlbom2022}, energy balance \cite{Garanin2021}, and symmetric splittings \cite{Ma2016,Tranchida2018,Ableidinger2016}.
To our knowledge, no method keeps LLG dynamics at one model evaluation per step and measures the spin temperature exactly for any Hamiltonian at no extra model cost.

\begin{figure*}[t]
  \centering
  \includegraphics[width=\textwidth]{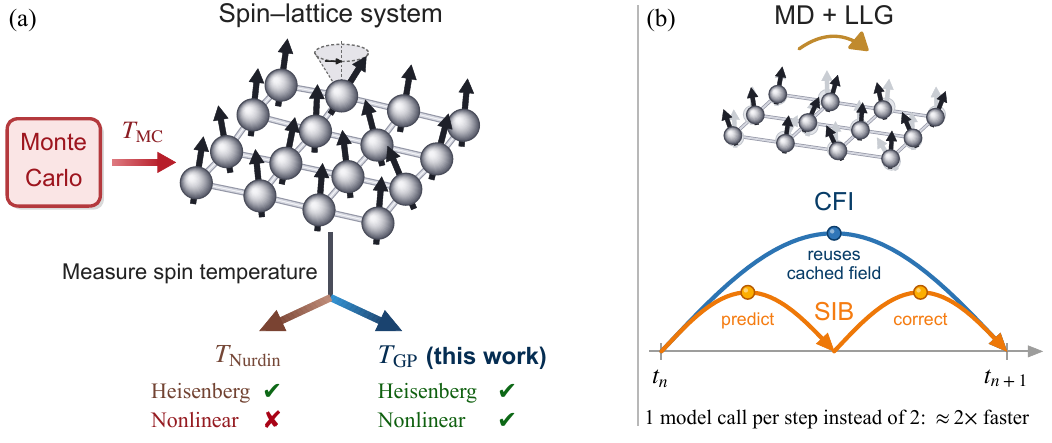}
  \caption{Spin thermometry and single-call SLD integration.
  (a) Monte Carlo fixes $T_{\mathrm{MC}}$. The simplified Nurdin reading $T_{\mathrm{Nurdin}}$ is exact for Heisenberg models, whose energy is linear in each spin, and fails for nonlinear Hamiltonians; the geometric-probe spin temperature (GPST) $T_{\mathrm{GP}}$ (this work) is computed from the spin forces alone and is exact for any smooth Hamiltonian.
  (b) One coupled MD + LLG step $t_n\to t_{n+1}$: SIB calls the model twice per step, to predict and to correct, whereas CFI reuses the cached field for its predictor and calls the model once. ``$\approx 2\times$ faster'' refers to the measured CFI/SIB per-step time ratio of $0.51$--$0.54$ on the SpinGNN potential.}
  \label{fig:schematic}
\end{figure*}

Here we keep LLG dynamics and address both problems (Fig.~\ref{fig:schematic}).
For the first, we obtain the spin temperature from the spin forces alone: the only Hamiltonian-dependent input is the spin force $\mathbf{f}_i=-\partial H/\partial\mathbf{s}_i$, which the integrator already computes, and the estimator is exact for any smooth Hamiltonian, whereas the simplified Nurdin reading, built from the same inputs, is exact only for energies linear in each spin.
We call this estimator the geometric-probe spin temperature (GPST), $T_{\mathrm{GP}}$, because it is built from geometric probe fields that we prescribe and that do not contain $H$.
For the second, the cached-field integrator (CFI) calls the model once per step instead of twice and remains second-order accurate.
We validate $T_{\mathrm{GP}}$ against independent Monte Carlo (MC) sampling on analytic bcc Fe models and on a SpinGNN potential, and compare CFI with SIB on Fe and on a SpinGNN potential for the layered antiferromagnet CrSBr \cite{Telford2020,Yu2024}.

\emph{Spin temperature from spin forces.}---We obtain the spin temperature from the spin forces alone, and the result is exact for any smooth Hamiltonian $H(\mathbf{x},\mathbf{s})$.
Let $\rho\propto e^{-\beta H}d\mathbf{x}\prod_i d\Omega(\mathbf{s}_i)$ with $|\mathbf{s}_i|=1$ and $\beta=1/k_BT$, and let $\mathbf{X}_i$ be any smooth field tangent to the unit sphere of spin $i$.
The argument holds at fixed positions $\mathbf{x}$, so it applies to frozen and moving lattices and to NPT, using the geometry of each frame.
Because the sphere has no boundary, $\oint\nabla_{S_i}\!\cdot(\mathbf{X}_i e^{-\beta H})\,d\Omega_i=0$, so
\begin{equation}
  \langle\nabla_{S_i}\!\cdot\mathbf{X}_i\rangle
  =\beta\langle\mathbf{X}_i\!\cdot\!\nabla_{S_i}H\rangle
  =-\beta\langle\mathbf{X}_i\!\cdot\!\mathbf{f}_i^{\perp}\rangle,
  \label{eq:ibp}
\end{equation}
where $\mathbf{f}_i^{\perp}=(\mathbf{I}-\mathbf{s}_i\mathbf{s}_i^{\mathsf{T}})\mathbf{f}_i=-\nabla_{S_i}H$ is the tangential part of the spin force.
Summing over spins gives
\begin{equation}
  \begin{aligned}
  &\langle N_X\rangle=k_BT\,\langle D_X\rangle,\\
  &N_X=\sum_i\mathbf{f}_i^{\perp}\!\cdot\mathbf{X}_i,\qquad
  D_X=-\sum_i\nabla_{S_i}\!\cdot\mathbf{X}_i .
  \end{aligned}
  \label{eq:identity}
\end{equation}
Equation~(\ref{eq:identity}) is known \cite{Rugh1997,Jepps2000,Davis2012}; for spins it is the $H$-independent special case of Eq.~(20) in Ref.~\cite{Gutierrez2018}, which holds for an arbitrary field.
That work built Rickayzen-type estimators \cite{Rickayzen2001}, whose divergence involves second derivatives of $H$ in general, gave explicit formulas only for the Heisenberg model, and reported no numerical tests.

The choice of $\mathbf{X}_i$ decides which derivatives of $H$ are needed: $H$ enters $N_X$ only through $\mathbf{f}_i^{\perp}$ but enters $D_X$ through $\nabla_{S_i}\!\cdot\mathbf{X}_i$.
The Nurdin choice $\mathbf{X}_i=\mathbf{f}_i^{\perp}$ gives $N=\sum_i|\mathbf{s}_i\times\mathbf{f}_i|^2$ and $D=\sum_i\Delta_{S_i}H$, a second derivative of $H$.
Only if $H$ is linear in $\mathbf{s}_i$ does $\Delta_{S_i}H$ reduce to $2\,\mathbf{s}_i\!\cdot\mathbf{f}_i$, which yields the simplified reading $T_{\mathrm{N}}=\langle\sum_i|\mathbf{s}_i\times\mathbf{f}_i|^2\rangle/(2k_B\langle\sum_i\mathbf{s}_i\!\cdot\mathbf{f}_i\rangle)$ \cite{Nurdin2000,Ma2010}.
For the GPST we instead take $\mathbf{X}_i=\nabla_{S_i}\Phi$ with a probe $\Phi(\mathbf{s};\mathbf{x})$: a scalar function of the spin directions and the atomic geometry that we prescribe and that does not contain $H$.
Then $D_X=-\sum_i\Delta_{S_i}\Phi$ is an explicit function of the configuration and $H$ enters only through $\mathbf{f}_i^{\perp}$: for any smooth Hamiltonian, $T_{\mathrm{GP}}$ follows from the spin forces alone.
Because only the tangential force enters, $T_{\mathrm{GP}}$ does not depend on how $H$ is continued off the unit spheres, whereas the denominator $\mathbf{s}_i\!\cdot\mathbf{f}_i$ of $T_{\mathrm{N}}$ is a radial derivative and depends on that continuation.
Our main probes are Legendre pair probes $\Phi_\ell=\sum_{i<j}w_{ij}P_\ell(c_{ij})$, where $c_{ij}=\mathbf{s}_i\!\cdot\mathbf{s}_j$, the geometric weights $w_{ij}$ select neighbor shells, and $\ell\in\{1,2,4,8\}$. Since $P_\ell$ is a degree-$\ell$ spherical harmonic of each spin, $\Delta_{S_i}P_\ell=-\ell(\ell+1)P_\ell$ and
\begin{equation}
  \begin{aligned}
  N_\ell&=\sum_{i<j}w_{ij}P'_\ell(c_{ij})\,(\mathbf{s}_j\!\cdot\mathbf{f}_i^{\perp}+\mathbf{s}_i\!\cdot\mathbf{f}_j^{\perp}),\\
  D_\ell&=2\ell(\ell+1)\sum_{i<j}w_{ij}P_\ell(c_{ij}).
  \end{aligned}
  \label{eq:legendre}
\end{equation}
Single-site quadratic, bond-direction, and chirality probes complete the set \cite{SM}.
We estimate $T_{\mathrm{GP}}=\bar N/(k_B\bar D)$ from ratios of ensemble means and combine the probes by a cross-fitted generalized method of moments, with whole-chain jackknife uncertainties and Fieller confidence sets \cite{Fieller1954,SM}. This requires $\langle D_X\rangle\neq0$ for at least one probe.
Forces stored with each configuration give $T_{\mathrm{GP}}$ without any extra model evaluation.
By comparison, a finite-difference configurational temperature with $8$ probes needs $17$ model calls per frame, one unperturbed and two perturbed evaluations per probe. A stochastic trace estimate of $\sum_i\Delta_{S_i}H$ \cite{Hutchinson1990} still needs at least one extra force evaluation per random vector, or a Hessian--vector product, and it adds sampling noise.
The probe sums cost $66.5$--$510.9~\mu$s per frame for $128$--$1024$ atoms in a CPU benchmark without output.

\begin{figure*}[t]
  \centering
  \includegraphics[width=0.85\textwidth]{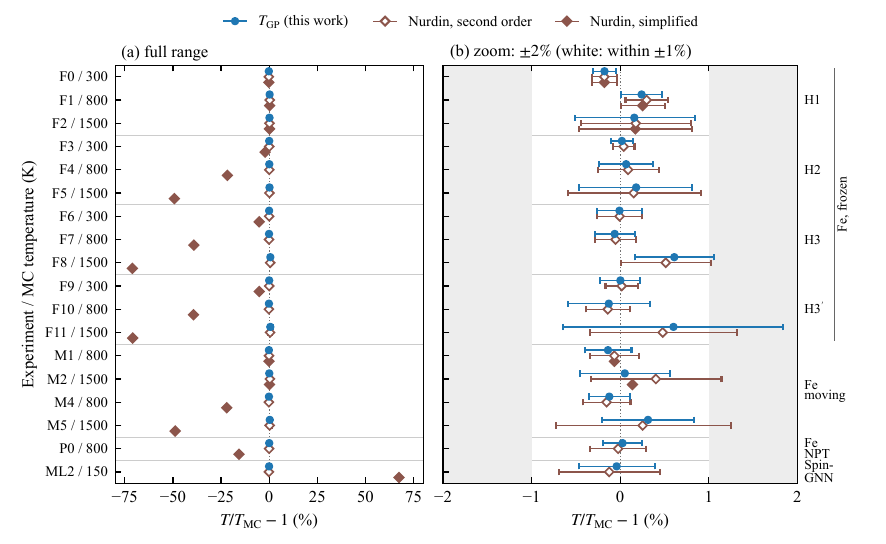}
  \caption{Spin thermometers compared with Monte Carlo. For each of the $18$ benchmarks, labeled by experiment and MC temperature, the relative deviation $T/T_{\mathrm{MC}}-1$ from the temperature $T_{\mathrm{MC}}$ set in MC is shown for the GPST $T_{\mathrm{GP}}$ (this work, blue circles), for the second-order Nurdin reading, which uses second derivatives of $H$ (open diamonds), and for the simplified Nurdin reading (filled diamonds).
  (a) Full range on a linear axis. (b) Enlarged view of $\pm2\%$ (white: within $\pm1\%$); the simplified Nurdin reading appears only where it falls in this range.
  F0--F11, frozen Fe (F0--F2 $H_1$, F3--F5 $H_2$, F6--F8 $H_3$, F9--F11 $H_3'$, i.e., $H_3$ on a perturbed geometry); M1, M2 ($K=0$) and M4, M5 ($K=0.02$~eV), moving Fe; P0, Fe NPT; ML2, second SpinGNN potential.
  Bars: ordinary $95\%$ Fieller intervals over $8$ chains; agreement is judged with pre-registered family-wise intervals. Simplified Nurdin points for M1--M5 and ML2 are point values without intervals; the second-order Nurdin reading for ML2 uses every fifth frame.}
  \label{fig:benchmark}
\end{figure*}

\emph{Cached-field integrator.}---SIB \cite{Mentink2010} advances each spin with a norm-preserving Cayley map in two stages \cite{SM}.
A predictor uses the spin field at the current spins $\mathbf{s}_n$ to estimate the midpoint $\mathbf{m}=(\mathbf{s}_n+\mathbf{p})/|\mathbf{s}_n+\mathbf{p}|$, and a corrector advances $\mathbf{s}_n$ with the field at $\mathbf{m}$; each field needs a model evaluation.
In the moving-lattice SIB used here as the baseline, the second evaluation is the velocity-Verlet force call at the new positions, so a step calls the model twice, once for the spin update and once for the atomic forces, and the spins are advanced at the old positions $\mathbf{x}_n$.

CFI rests on one observation: the predictor only has to locate the midpoint.
Its field enters the spin update multiplied by the time step $h$, so an error of order $h$ in that field moves $\mathbf{m}$ by order $h^2$, and the corrector, evaluated with the freshly computed field at $\mathbf{m}$, still yields a second-order step.
CFI therefore reuses the field from the previous step's midpoint evaluation, which differs from the field at $\mathbf{s}_n$ by order $h$, instead of calling the model at $\mathbf{s}_n$.
To let this single evaluation serve both the spins and the lattice, CFI orders each step symmetrically: (1) half a lattice Ornstein--Uhlenbeck (OU) bath step and a half drift bring the atoms to $\mathbf{x}_{n+1/2}$; (2) the cached field gives the predictor and the midpoint $\mathbf{m}$; (3) one model evaluation at $(\mathbf{x}_{n+1/2},\mathbf{m})$ returns the atomic forces, the spin field, and, in NPT, the virial; (4) the velocities receive a full kick, the spins are corrected with the new field, and the field is cached; (5) a half drift and half an OU step complete the step.
Each step thus calls the model once [Fig.~\ref{fig:schematic}(b)].
When a configuration is written out, one additional evaluation at the end-of-step state supplies forces that match the stored configuration; it does not enter the dynamics, and its cost scales with the number of written frames, not with the number of steps \cite{SM}.
The symmetric order gives second-order accuracy on a moving lattice, and the cached-field predictor halves the evaluations; an ablation with the same order and a fresh predictor separates the two \cite{SM}.
With baths off and $\alpha=0$, CFI converges with $\mathbf{x}/\mathbf{v}/\mathbf{s}$ orders $1.9991$--$2.0002$ and energy order $1.9994$--$2.0008$ on a $4$-atom system and $1.9955$--$1.9998$ in bulk Fe, whereas the SIB baseline has spin order $1.0022$--$1.0118$ on the $4$-atom system and a short-window spin slope of $1.91$--$1.94$ in bulk Fe.

\emph{Validation against Monte Carlo.}---Figure~\ref{fig:benchmark} compares $T_{\mathrm{GP}}$ with the MC temperature $T_{\mathrm{MC}}$ for $18$ benchmarks \cite{SM}: frozen-lattice bcc Fe ($128$ atoms) with isotropic exchange $H_1$, linear in each spin, with $H_2$, adding uniaxial single-ion anisotropy, and with $H_3$, adding a Dzyaloshinskii--Moriya interaction (DMI) and biquadratic exchange; moving-lattice Fe NVT with single-ion anisotropy $K=0$ or $0.02$~eV; Fe NPT; and, at $150$~K, a second SpinGNN potential for CrSBr ($108$ atoms, $36$ Cr spins), trained independently of the one used for the dynamics below and evaluated in a deterministic precision mode so that MC acceptance is reproducible.
Every MC reference passed its sampling-qualification gates \cite{Flyvbjerg1989,Gelman1992}.

$T_{\mathrm{GP}}$ agrees with $T_{\mathrm{MC}}$ within $0.62\%$ in all $18$ benchmarks, for linear, nonlinear, and ML Hamiltonians and for frozen, moving, and constant-pressure lattices.
The simplified Nurdin reading, shown for all $18$ benchmarks including $H_3$ on a perturbed geometry ($H_3'$), is accurate only for the linear Hamiltonians ($H_1$ and moving Fe with $K=0$); it deviates by up to $-71\%$ for the nonlinear analytic models and by $+67\%$ for the SpinGNN potential. Its bias therefore comes from nonlinearity, as the derivation of Eq.~(\ref{eq:identity}) shows.
The second-order Nurdin reading, which uses $D=\sum_i\Delta_{S_i}H$ instead of $2\sum_i\mathbf{s}_i\!\cdot\mathbf{f}_i$, agrees with $T_{\mathrm{MC}}$ within $0.52\%$ in all $18$ benchmarks, but it needs $2$--$32$ extra force evaluations per frame for the analytic Fe models and $72$ Hessian--vector products per frame for the SpinGNN potential \cite{SM}; $T_{\mathrm{GP}}$ needs none.

\begin{figure}[t]
  \centering
  \includegraphics[width=\columnwidth]{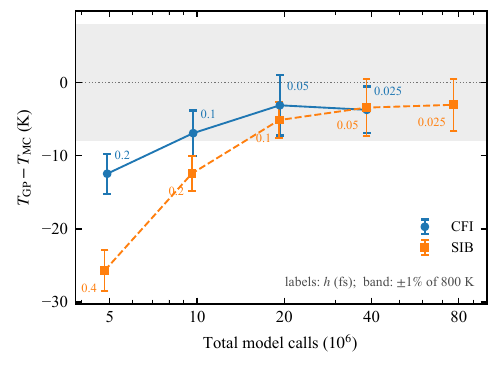}
  \caption{Accuracy versus cost of SIB and CFI. GPST bias $T_{\mathrm{GP}}-T_{\mathrm{MC}}$ versus total model evaluations (logarithmic axis) for moving-lattice Fe NVT at $800$~K ($128$ atoms, $K=0.02$~eV, $\mu=2.2\,\mu_B$, $\alpha=1$).
  Numbers give the step $h$ (fs): $0.2$--$0.025$ for CFI (blue circles) and $0.4$--$0.025$ for SIB (orange squares).
  Bars: ordinary $95\%$ confidence intervals over $8$ chains. Shaded band: $\pm1\%$ of $800$~K.}
  \label{fig:nvt}
\end{figure}

\begin{figure*}[t]
  \centering
  \includegraphics[width=0.95\textwidth]{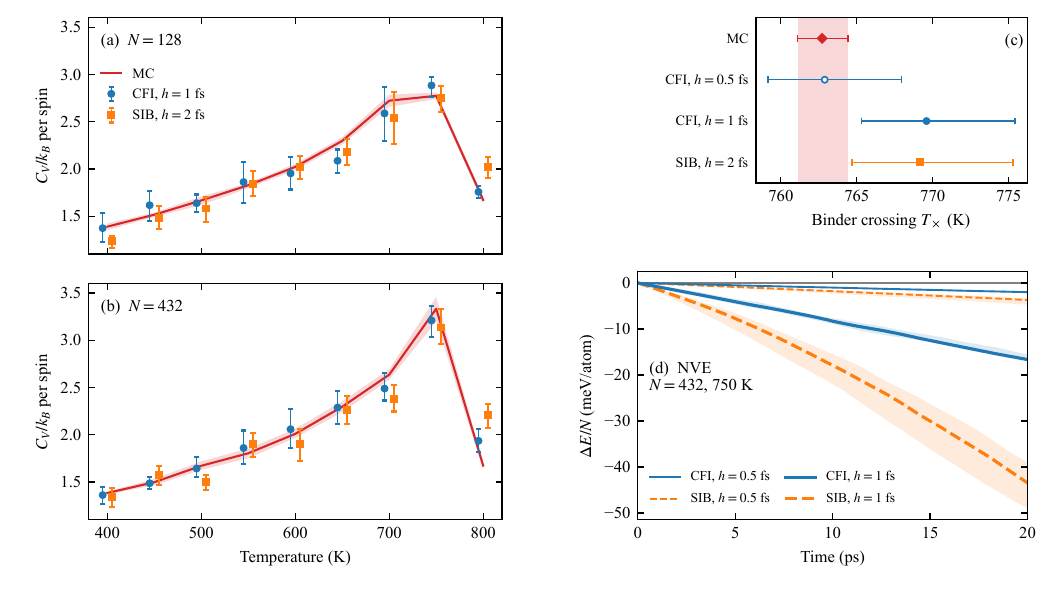}
  \caption{Thermodynamics and energy conservation of SIB and CFI on frozen-lattice bcc Fe ($H_3$). Specific heat $C_V/k_B$ per spin for (a) $N=128$ and (b) $N=432$: MC (red line and $95\%$ band; lines connect grid points), CFI at $h=1$~fs (blue circles), and SIB at $h=2$~fs (orange squares), which use the same number of model evaluations. CFI and SIB points are offset by $-5$ and $+5$~K.
  (c) Two-size Binder crossing temperature $T_\times$, a finite-size crossing rather than a bulk ordering temperature; the shaded band is the MC $95\%$ interval.
  (d) Energy drift $\Delta E/N$ in NVE (no baths, $\alpha=0$) for $N=432$ at $750$~K: CFI (solid) and SIB (dashed) at equal step $h$ ($0.5$ and $1$~fs) from the same initial states; lines are medians and bands the ranges over $8$ initial states.
  All bars are ordinary $95\%$ confidence intervals over $8$ chains.}
  \label{fig:thermo}
\end{figure*}

\emph{Comparison of SIB and CFI.}---Figure~\ref{fig:nvt} shows the GPST bias of moving-lattice Fe NVT at $800$~K against MC as a function of the total number of model evaluations.
At the same $h$ the two integrators have the same finite-step bias, and both are within $1\%$ of $T_{\mathrm{MC}}$ at $h=0.05$~fs; because CFI needs half the evaluations per step, its curve is shifted to half the number of evaluations, and its bias is smaller wherever the finite-step bias dominates.
Against an ablation with the same symmetric order and two evaluations per step, two one-sided equivalence tests \cite{Schuirmann1987} show equivalence at the same $h$ in all tested observables on Fe and on CrSBr \cite{SM}.
In deterministic NVE at equal model-call budget, CFI has smaller trajectory errors than SIB and, for $h\le 0.5$~fs, smaller long-run energy errors \cite{SM}.

Per step, CFI takes $0.598$--$0.613$ of the SIB wall-clock time for the analytic Fe potential and $0.509$--$0.541$ for the CrSBr SpinGNN potential, close to the evaluation ratio $1/2$ because model evaluation dominates the cost with an ML potential \cite{SM}.
On the CrSBr SpinGNN potential ($108$ atoms, $h=0.5$~fs), $T_{\mathrm{GP}}$ from both integrators stays within $1.6\%$ of the bath temperature at $80$--$120$~K \cite{SM}.
At a single NPT state point, Fe with an analytic potential at $800$~K, CFI agrees with independent NPT MC in volume, energy, magnetization, and $T_{\mathrm{GP}}$ within pre-registered equivalence margins \cite{SM}.

\emph{Thermodynamics.}---Figure~\ref{fig:thermo} tests the integrators across the magnetic transition of frozen-lattice Fe $H_3$ for $N=128$ and $432$ spins, with CFI at $h=1$~fs and SIB at $h=2$~fs, which use the same number of model evaluations.
Both dynamical results overlap the MC band at most temperatures, and the largest deviations occur at $800$~K, above the crossing [Fig.~\ref{fig:thermo}(a,b)].
The two-size Binder-cumulant crossing \cite{Binder1981} from CFI at $h=0.5$~fs agrees with MC within its $95\%$ interval; at equal model-call budget, CFI at $h=1$~fs and SIB at $h=2$~fs give the same crossing, both about $7$~K above MC [Fig.~\ref{fig:thermo}(c)], so removing one evaluation per step adds no resolvable bias.
In NVE without baths ($\alpha=0$, $N=432$, $750$~K), started from the same states at the same $h$, the median energy drift $\Delta E/N$ after $20$~ps is $-2.01$~meV/atom for CFI and $-3.72$~meV/atom for SIB at $h=0.5$~fs, and $-16.65$ and $-43.48$~meV/atom at $h=1$~fs [Fig.~\ref{fig:thermo}(d)].

\emph{Discussion.}---We obtain the spin temperature from the spin forces alone; the GPST is exact for any smooth Hamiltonian, needs no extra model evaluation, and agrees with MC within $0.62\%$ for linear, nonlinear, and ML Hamiltonians, where the simplified Nurdin reading errs by $-71\%$ to $+67\%$.
CFI keeps LLG dynamics at one model evaluation per step, has the same finite-step bias as SIB at the same step, and takes about half of its time per step with an ML potential.
The MC validation covers analytic Fe and one SpinGNN potential, because the energy evaluation of the CrSBr model used for the dynamics is not reproducible at the MC acceptance scale.

\begin{acknowledgments}
We acknowledge financial support from the National Key Research and Development Program of China (grant no.~2022YFA1402901, H.X.), NSFC (grant no.~12504262, Y.Z.), NSFC (grant no.~12188101, H.X.), Shanghai Science and Technology Program (grant no.~23JC1400900, H.X.), the Guangdong Major Project of the Basic and Applied Basic Research (Future functional materials under extreme conditions---grant no.~2021B0301030005, H.X.), Shanghai Pilot Research---Fudan University 21TQ1400100 (grant no.~23TQ017, H.X.), the robotic AI-Scientist platform of Chinese Academy of Sciences (H.X.) and New Cornerstone Science Foundation (H.X.).
\end{acknowledgments}

\emph{Code availability.}---The codes are made public at \url{https://github.com/Hyyu-Lab/spingnn-lammps}.

\bibliography{refs}

\end{document}